\documentclass[prb,twocolumn,showpacs,amsmath,amssymb]{revtex4}

\usepackage[pdftex]{graphicx}
\usepackage{enumerate}
\usepackage{longtable}

\begin{document}
\title{Temperature-dependent optical conductivity of layered LaSrFeO$_4$ }
\author{J. Reul, L. Fels, N. Qureshi, K. Shportko, M. Braden, and M. Gr{\"u}ninger}
\affiliation{~II.~Physikalisches Institut,
Universit{\"a}t zu K{\"o}ln, Z{\"u}lpicher Strasse 77, D-50937 K{\"o}ln, Germany}

\begin{abstract}
Compounds with intermediate-size transition metals such as Fe or Mn are close to the transition between
charge-transfer systems and Mott-Hubbard systems.
We study the optical conductivity $\sigma(\omega)$ of insulating layered LaSrFeO$_4$ in the energy range
$0.5-5.5$\,eV from 15\,K to 250\,K by the use of spectroscopic ellipsometry in combination with transmittance measurements.
A multipeak structure is observed in both $\sigma^{a}(\omega)$ and $\sigma^c(\omega)$.
The layered structure gives rise to a pronounced anisotropy,
thereby offering a means to disentangle Mott-Hubbard and charge-transfer absorption bands.
We find strong evidence that the lowest dipole-allowed excitation in LaSrFeO$_4$ is of Mott-Hubbard type.
This rather unexpected result can be attributed to Fe $3d$ - O $2p$ hybridization and in particular to the
layered structure with the associated splitting of the $e_g$ level.
In general, Mott-Hubbard absorption bands may show a strong dependence on temperature.
This is not the case in LaSrFeO$_4$, in agreement with the fact that spin-spin and orbital-orbital correlations
between nearest neighbors do not vary strongly below room temperature in this compound with a
high-spin $3d^5$ configuration and a N\'{e}el temperature of $T_N=366$\,K.
\end{abstract}


% PACS-NUMBERS
% 71.        Electronic structure of bulk materials
% 71.20.Be   Transition metals and alloys
% 71.27.+a   Strongly correlated electron systems; heavy fermions
% 78.        Optical properties, condensed-matter spectroscopy and
%            other interactions of radiation and particles with condensed matter
% 78.20.-e   Optical properties of bulk materials and thin films
% 78.20.Ci   Optical constants
% 78.66.Nk   Insulators

\date{March 4, 2013}
\maketitle

\section{Introduction}

\begin{figure}[b]
\includegraphics[width=0.5\columnwidth,angle=270,clip]{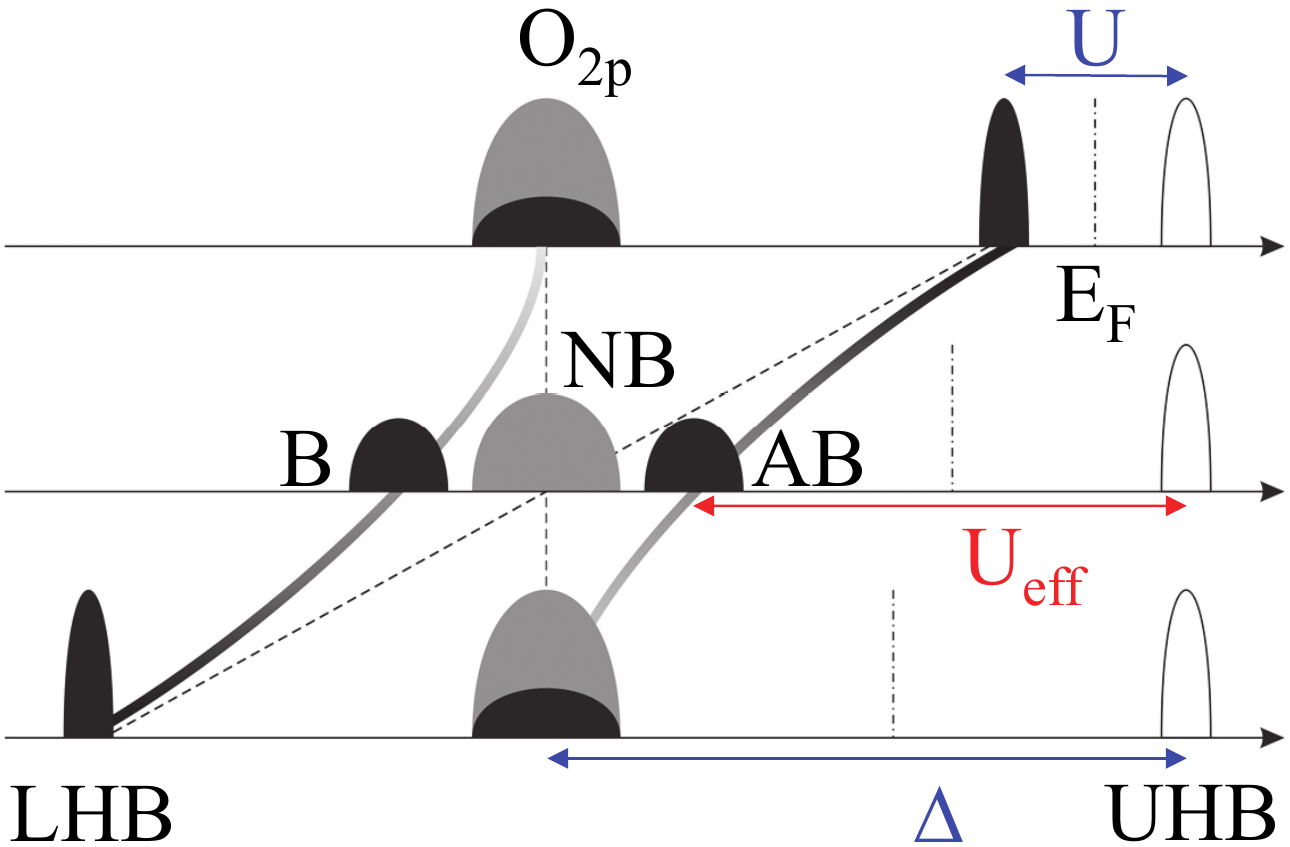}
\caption{(Color online) Sketch of a Mott-Hubbard insulator (top row, $U \ll \Delta$) and of a charge-transfer insulator
(bottom row, $U \gg \Delta$) for a single half-filled $3d$ orbital and degenerate O $2p$ orbitals.
The black dashed line depicts the increase of $U$ from top to bottom, $E_F$ depicts the Fermi level (dash-dotted).
Due to hybridization, we have to distinguish bonding (B), non-bonding (NB), and anti-bonding (AB) bands.
For $U \gtrsim \Delta$, the highest occupied anti-bonding band may still be classified as the lower Hubbard band
with symmetry properties derived from the $3d$ character, yielding an effective value $U_{\rm eff} < \Delta$
(cf.\ Fig.\ 1 in Ref.\ [\onlinecite{goessling08}] and Fig.\ 2 in Ref.\ [\onlinecite{olalde}]).  }
\label{MHCT}
\end{figure}

Many transition-metal compounds with a partially filled $3d$ shell show charge localization and insulating behavior.
Typically, this is driven by the large on-site Coulomb repulsion $U$, which splits the conduction band into a lower and an upper Hubbard
band (LHB and UHB).\@ According to the  Zaanen-Sawatzky-Allen scheme,\cite{ZSA} one distinguishes between two kinds of correlated
insulators, depending on the relative size of $U$ and the charge-transfer (CT) energy $\Delta$ between the transition-metal $3d$ band
and the highest occupied ligand level, e.g.\ oxygen $2p$. In a CT insulator with $U > \Delta$, the charge gap is formed between
O $2p$ and the UHB.\@ In contrast, Mott-Hubbard (MH) insulators show $U < \Delta$, and the states closest to the Fermi level
predominantly have transition-metal character (see Fig.\ \ref{MHCT}).
This character is decisive for a quantitative description of doped compounds.
For instance in a hole-doped CT system such as the high-$T_c$ cuprates, the mobile carriers predominantly have oxygen character.

Early transition-metal compounds are typically classified into the group of MH insulators, whereas late
ones are identified as being of CT type.\cite{zaanen90,mizokawa1996,matsuno,olalde,arima1993}
With increasing atomic number, $U$ increases whereas $\Delta$ decreases,\cite{zaanen90,arima1993}
which reflects the decreasing spatial extension of the $3d$ orbitals and the increasing electronegativity
of the transition-metal ions, respectively.
However, the character of oxides with intermediate size transition-metal ions, in particular Mn$^{3+}$, has been
discussed controversially. On the basis of optical data, both LaMnO$_3$ and LaSrMnO$_4$ have been interpreted either
as of CT type\cite{arima1993,arima1995,tobe2001,moritomo1995,lee07}
or of MH type,\cite{jung1995,quijada2001,kovaleva2004a,kim04,kim06,goessling08}
while recently a dual nature of the optical gap has been proposed.\cite{moskvin}
This controversy arises due to the strong hybridization between Mn $3d$ and O $2p$ states (see Fig.\ \ref{MHCT}).
Early on, Mizokawa and Fujimori\cite{mizokawa1996} pointed out that $U > \Delta$ in LaMnO$_3$, but that the
highest occupied O $2p$ band shows a large admixture of $3d$ character.
Goessling \textit{et al.}\cite{goessling08} emphasized that the symmetry of the highest occupied, strongly hybridized band
is determined by the $3d$ band, which is essential for the optical selection rules and thus for a quantitative analysis of
the optical data. They suggested that the manganites can be viewed as \textit{effective} Mott-Hubbard systems,
where $U_{\rm eff}$ is strongly renormalized by hybridization (see Fig.\ \ref{MHCT}).
This scenario is supported by recent measurements on transition-metal difluorides $M$F$_2$ using x-ray emission
spectroscopy.\cite{olalde} Due to the element selectivity of this technique, the contribution of the transition-metal LHB
to the highest occupied states can be revealed even for $U > \Delta$.

At first sight, the situation is more transparent in the case of the ferrites $R$FeO$_3$ with trivalent Fe ions.
Compared to the $3d^4$ manganites, the stability of the high-spin $3d^5$ state of Fe$^{3+}$ gives rise to a
comparably large energy of MH excitations of roughly $U + 4 J_H$, where $J_H$ denotes the intra-atomic Hund exchange.\@
Indeed these compounds commonly are identified as CT systems.\cite{arima1993,arima1995,mizokawa1996,pisarev09}
However, the case of layered LaSrFeO$_4$ is still under discussion. It shows the same Fe$^{3+}$ $3d^5$ configuration
and has also been interpreted as a CT insulator based on optical reflectivity data measured up to 36\,eV at room-temperature
with in-plane polarization of the electric field.\cite{moritomo1995}
In contrast, Omata \textit{et al.} \cite{omata1994XAS} conclude from their resonant photoemission data
that the valence band in LaSrFeO$_4$ is formed by a mixture of Fe $3d$ and O $2p$ states.
Thus they characterize LaSrFeO$_4$ as an intermediate type CT and MH insulator, but they also mention that
the states at the valence band edge mainly show O $2p$ character.
Here, we address the optical conductivity $\sigma_1(\omega)$ of LaSrFeO$_4$. The anisotropy of $\sigma_1(\omega)$
of this layered structure provides the key to disentangle MH excitations and CT excitations.
Neglecting hybridization, a MH excitation refers to an electron transfer between neighboring Fe sites $i$ and $j$,
$|3d^n_i \, 3d^n_j\rangle \, \rightarrow \, |3d^{n+1}_i \, 3d^{n-1}_j\rangle$,
whereas a CT excitation refers to a transfer from O $2p$ to Fe $3d$,
$|3d^n \, 2p^6\rangle \, \rightarrow \, |3d^{n+1} \, 2p^{5}\rangle$.
In the presence of hybridization, this distinction between MH and CT excitations
is still a valid classification scheme since the hybridized states
retain their original symmetry and thus follow the same selection rules.
As discussed by Goessling \textit{et al.}\cite{goessling08} for the case of LaSrMnO$_4$, MH excitations
only contribute to $\sigma_1^a$, i.e.\ for polarization of the electric field within the 2D layer,
but not to $\sigma_1^c$. This reflects that Fe\,-\,Fe hopping between adjacent FeO$_2$ layers is negligible.
On the contrary, CT excitations are observed in both, $\sigma_1^a(\omega)$ and $\sigma_1^c(\omega)$,
as each Fe site is surrounded by an oxygen octahedron.
We find a clear anisotropy of the lowest dipole-allowed excitations in LaSrFeO$_4$, which suggests that they are
of Mott-Hubbard type. We propose that this is driven by Fe $3d$ - O $2p$ hybridization and by the layered structure,
the associated crystal-field splitting of the $e_g$ level pulls the lowest MH absorption band below the onset
of CT excitations.

Additionally, we address the temperature dependence of the spectral weight.
In general, CT and MH excitations show different spin and orbital selection rules.
The spectral weight of MH excitations is expected to be strongly affected by a change of the nearest-neighbor spin-spin
and orbital-orbital correlation functions.\cite{khaliullin2004a,khaliullinrev,oles2005,kovaleva2004a,goessling08,lee2005a,fang2003,goesslingTi,reul}
In agreement with theoretical expectations, the spectral weight of the lowest absorption band in LaMnO$_3$, LaSrMnO$_4$,
and $3d^2$ $R$VO$_3$ changes by a factor of 2 - 3 due to the ordering of spins (and orbitals).\cite{kovaleva2004a,goessling08,reul}
This clearly demonstrates the (\textit{effective}) MH character of these systems.
However, the behavior of the MH insulators YTiO$_3$ and SmTiO$_4$ is still puzzling in this context.\cite{goesslingTi}
In YTiO$_3$, the spectral weight of the lowest MH excitation is expected to change by 25\,\% between the paramagnetic
and the ferromagnetic state.\cite{oles2005} However, the increase around the ordering temperature $T_C$ amounts to only 5\,\%,
while, at the same time, larger changes are observed up to 300\,K.\cite{goesslingTi}
Due to the three-dimensional character of the magnetic order, these larger changes far above $T_C$ can not be explained
by a change of the spin-spin correlations.
In the G-type antiferromagnet SmTiO$_3$, spin ordering is expected to suppress the spectral weight of the lowest MH excitation
by about 50\,\% for all crystallographic directions,\cite{oles2005} but the observed effects are again much smaller and
show even the wrong sign along $b$ and $c$.\cite{goesslingTi}
This behavior has been attributed to small changes of the orbital occupation.\cite{goesslingTi}
It is important to quantify the possible strength of other effects such as excitonic contributions,
the thermal expansion of the lattice or bandstructure effects.
In this context, layered LaSrFeO$_4$ with its stable $3d^5$ state is an interesting candidate for a reference system.
Long-range antiferromagnetic spin order sets in at $T_N = 366$\,K,\cite{soubeyroux,omata94TN,qureshi} thus one expects
only very small changes of the spin-spin and orbital-orbital correlation functions below room temperature.
Thus far, not much is known about the optical spectra of LaSrFeO$_4$.
Room temperature data has been obtained by the means of diffuse reflectance on powdered samples\cite{omata1996optics}
and by reflectivity measurements on single crystals,\cite{moritomo1995,tajima} but only for
in-plane polarization of the electric field.

The paper is organized as follows. The experimental details are described in Sec.\ \ref{experiment},
followed in Sec.\ \ref{struc} by a short description of the crystal structure and the electronic structure.
A discussion of the expected multiplet splitting of MH and CT absorption bands is given in Sec.\ \ref{possex}.
Finally, in Sec.\ \ref{results} we present our experimental data (Sec.\ \ref{expdata}) together with a discussion
of the peak assignment (Sec.\ \ref{peakassignment}) and an analysis of the temperature dependence of
the MH excitations (Sec.\ \ref{tdep}).

\section{Experiment}
\label{experiment}

\begin{figure}[tb]
\includegraphics[width=0.85\columnwidth,clip]{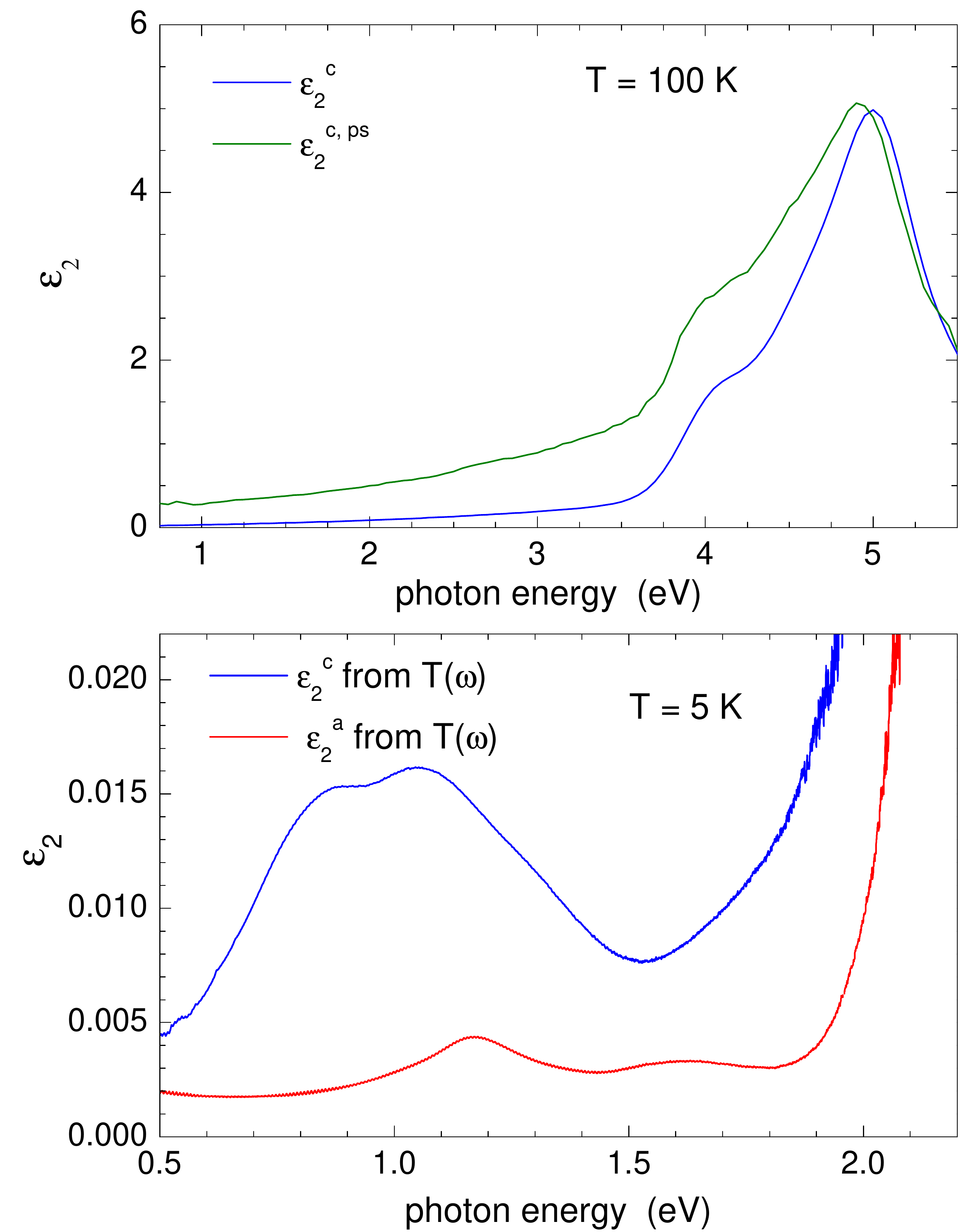}
\caption{(Color online) Top: Comparison of the dielectric function $\varepsilon_2^{c}$ and the pseudo-dielectric
function $\varepsilon_2^{c,ps}$. For the former, a surface roughness of 6\,nm has been taken into account.
Bottom: $\varepsilon_2(\omega)$ below the charge gap as determined from transmittance measurements on a single crystal
with a thickness of 39\,$\mu$m. Note the much smaller scale of $\varepsilon_2$ in the bottom panel.
}
\label{pseudo}
\end{figure}

Single crystals of LaSrFeO$_4$ have been grown using the floating-zone method.\cite{qureshi}
The purity, stoichiometry, and single-phase structure of the crystals was checked by neutron diffraction and by x-ray diffraction.
Typical dimensions of crystals used in this study are a few mm along all three crystallographic axes.
Ellipsometric data in the energy range $0.75-5.5$ eV was obtained using a rotating-analyzer ellipsometer (Woollam VASE)
equipped with a retarder between polarizer and sample. The angle of incidence was $\theta = 70^\circ$.
The ellipsometric measurements have been performed from 15\,K to 250\,K in a UHV cryostat with $p < 10^{-9}$\,mbar.
Window effects of the cryostat have been corrected using a standard Si wafer.
Ellipsometry is a self-normalizing technique and does not require reference measurements, furthermore it
yields the complex dielectric function directly without a Kramers-Kronig transformation. These are two significant advantages
over conventional reflection measurements.
Ellipsometry is particularly suited to determine the precise temperature dependence of the
optical spectral weight.\cite{kovaleva2004a,goessling08,goesslingTi,reul}
Measurements of the ellipsometric angles $\Psi$ and $\Delta$ have been performed on a polished $ac$ surface in
two different orientations, with the $a\ (c)$ axis perpendicular to the plane of incidence.
We obtain the two non-vanishing, complex entries $\varepsilon^{a}$ and $\varepsilon^{c}$ of the dielectric tensor for
tetragonal symmetry by fitting the measured data of both orientations simultaneously
with a series of Gaussian oscillators for $\varepsilon_2$ and a Kramers-Kronig consistent line shape for $\varepsilon_1$.
In the analysis, we have taken into account a finite surface roughness since ellipsometry is a surface-sensitive technique.
The properties of the surface can be determined reliably in a frequency range where the investigated bulk sample is transparent,
i.e., $\varepsilon_2(\omega) \approx 0$.
To determine the suitable energy range, we employed a Fourier-transform spectrometer (Bruker IFS 66/v)
and performed infrared transmittance measurements between 0.5\,eV and 2.5\,eV on a single crystal polished to a thickness
of 39\,$\mu$m. We used the observed interference fringes to determine the refractive index $n$, which in turn allowed us to
determine $\varepsilon_2(\omega)$ from the transmittance, see bottom panel of Fig.\ \ref{pseudo}.
These data show only very weak parity- and spin-forbidden local crystal-field excitations $3d^5 \rightarrow 3d^{5,*}$
between 0.5\,eV and 2.0\,eV, thus $\varepsilon_2(\omega) \approx 0$ is a valid approximation below 2\,eV.\@
Using this result for the analysis of the low-energy ellipsometry data yields a surface roughness with a thickness
of approx.\ 6\,nm.

In the top panel of Fig.\ \ref{pseudo} we compare our result for $\varepsilon_2^c(\omega)$ with the so-called
pseudo-dielectric function $\varepsilon_2^{c,ps}$.
Aspnes\cite{aspnes} proposed that the latter may serve as a reasonable approximation for the former under certain conditions.
The main advantage of the pseudo-dielectric function is that it can be determined directly from the data measured for a
\textit{single} orientation. Figure \ref{pseudo} shows that the overall features are well reproduced, but large discrepancies
are observed below the onset of strong absorption at about 3.5\,eV, where $\varepsilon_2^{c,ps}$ shows a spurious background.
As far as Fe$^{3+}$ compounds are concerned, a similar feature has been discussed in BiFeO$_3$.\cite{pisarev09}
In the case of LaSrFeO$_4$, the transmittance data prove that this background is an artefact present only in
$\varepsilon_2^{c,ps}$ but not in $\varepsilon_2^{c}$.

\section{Structure, spins, and orbitals}
\label{struc}

The compound LaSrFeO$_4$ crystallizes in the single-layered structure of K$_2$NiF$_4$ with tetragonal symmetry
\textit{I4/mmm}. The Fe$^{3+}$ ions are octahedrally coordinated by oxygen
ions, building perfect FeO$_2$ square planes with 180$^\circ$ Fe-O-Fe bonds.
The lattice constants at room temperature are $a= 3.8744(1)$ \AA \ and $c=12.7134(3)$ \AA.\cite{qureshi}
Nominally, there are five electrons in the $3d$ shell per Fe site.
In the high-spin ground state, these five electrons yield a total spin of 5/2.
Antiferromagnetic order has been observed below $T_N=366$\,K.\cite{soubeyroux,qureshi}
Our crystals do not exhibit any evidence for an additional magnetic phase transition.\cite{qureshi}
In cubic approximation, the $3d$ level is split into the lower-lying $t_{2g}$ and the higher-lying $e_g$ levels.
The magnitude of the $t_{2g} - e_g$ splitting $\Delta_{t2g-eg}=10$\,Dq is mainly determined
by the Fe - O bond lengths, it can be estimated to be roughly 10\,Dq\,=\,1.0\,-\,1.5\,eV
in LaSrFeO$_4$.\cite{goessling08,galuza1998}
The FeO$_6$ octahedra show a sizeable tetragonal distortion with Fe\,-\,O bond lengths of 1.9354\,\AA \
in the plane and 2.1486\, \AA \ perpendicular to it at 10\,K.\cite{qureshi}
Therefore, the $t_{2g}$ manifold is split into the lower-lying doublet $e_g^\prime$ ($yz$ and $zx$)
and the higher-lying $b_{2g}$ (or $xy$) level. At the same time, the $e_g$ level splits into
$a_{1g}$ ($3z^2 \! - \! r^2$) and $b_{1g}$ ($x^2 \! - \! y^2$), where the energy of the former
is significantly reduced compared to the energy of the latter.
In LaSrMnO$_4$, these splittings have been determined from optical data, yielding $\Delta_{t2g}(d^4)=0.2$\,eV
and $\Delta_{eg}(d^4)=1.4$\,eV.\cite{goessling08} For LaSrFeO$_4$, we expect a similar value of $\Delta_{t2g}$
but a smaller value of $\Delta_{eg}$, since $\Delta_{eg}$ is enhanced in LaSrMnO$_4$ due to the
additional Jahn-Teller splitting of the singly occupied $e_g$ level, which is evident from the larger value of $c/a$.
We emphasize that the pronounced elongation of the octahedra in LaSrFeO$_4$ gives rise to a sizeable value of
$\Delta_{eg}$, even in the absence of a Jahn-Teller contribution. Our results below yield
$\Delta_{eg} \approx 0.8$\,eV.\@

\section {Charge-transfer and Mott-Hubbard excitations}
\label{possex}

First, we focus on the physics of CT excitations. These result from the transfer of an electron from a
ligand O $2p$ orbital into a Fe $3d$ orbital,
% $|$Fe $3d^5$ O $2p^6\rangle \rightarrow |$Fe $3d^6$ O $2p^5\rangle$.
$|3d^5 \, 2p^6\rangle \, \rightarrow \, |3d^6 \, 2p^5\rangle$.
Pisarev \textit{et al.}\cite{pisarev09} present a detailed theoretical analysis of the CT excitations
for undistorted FeO$_6$ octahedra.
The highest occupied O states are of non-bonding character with symmetry $t_{1g}(\pi)$, $t_{2u}(\pi)$, $t_{1u}(\pi)$,
and $t_{1u}(\sigma)$.
Their relative energies are determined by, e.g., the different Madelung energies of $2p(\pi)$ and $2p(\sigma)$ orbitals and
by the $2p(\pi)$ - $2p(\pi)$ overlap.\cite{pisarev09} The $t_{1g}(\pi)$ level is expected to be the highest in energy.
Quantum-chemistry calculations\cite{pisarev09} for LaFeO$_3$ predict that
$t_{2u}(\pi)$, $t_{1u}(\pi)$, and $t_{1u}(\sigma)$ are lower by 0.8\,eV, 1.8\,eV, and 3\,eV, respectively.
The lowest unoccupied states are the anti-bonding $t_{2g}(\pi)$ and $e_g(\sigma)$ orbitals with hybrid Fe $3d$ - O $2p$ character,
and these are split by $\Delta_{t2g-eg}$\,=\,10\,Dq.

According to the parity selection rule, the even-even (from $g$ type to $g$ type) transitions from the $t_{1g}(\pi)$ level
at the top of the O band to the unoccupied $t_{2g}(\pi)$ and $e_g(\sigma)$ orbitals are forbidden.
Additionally, the matrix elements for transitions from $\pi$ to $\sigma$ levels vanish for a single octahedron,
thus only $\pi - \pi$ and $\sigma - \sigma$ transitions give rise to strong absorption.
In summary, the onset of CT excitations is governed by the dipole-forbidden transition $t_{1g}(\pi)\rightarrow t_{2g}(\pi)$,
followed by the strong dipole-allowed transitions $t_{2u}(\pi)\rightarrow t_{2g}(\pi)$ and $t_{1u}(\pi)\rightarrow t_{2g}(\pi)$.
The next transition $t_{1u}(\sigma)\rightarrow e_{g}(\sigma)$ is roughly 2\,eV higher in energy due to the splitting
between $t_{2g}(\pi)$ and $e_g(\sigma)$ on the one hand and between $2p(\pi)$ and $2p(\sigma)$ states on the other hand.
Thus in cubic approximation there are only two strong excitations in the energy range relevant to us.
For the layered structure of LaSrFeO$_4$, we have to expect
additional splittings. However, the large splitting $\Delta_{eg}$ of the $e_g$ level is not important for the lowest CT excitations,
which correspond to an electron transfer into the $t_{2g}(\pi)$ level.

\begin{figure}[tb]
\includegraphics[width=1.0\columnwidth,clip]{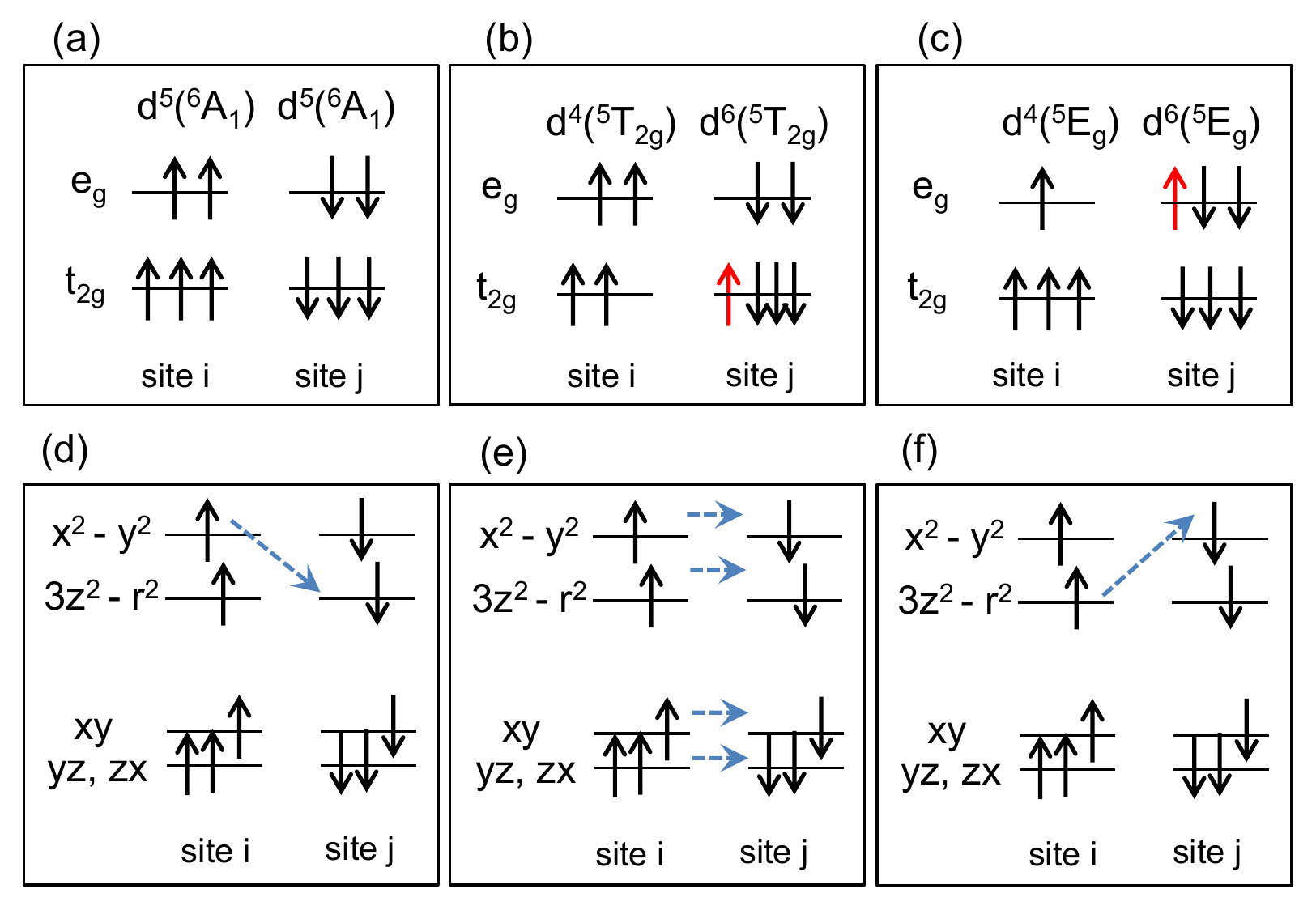}
\caption{(Color online)
Sketch of (a) the initial state $|3d^5_i \, 3d^5_j\rangle$ and
(b,c) the final states $|3d^4_i \, 3d^6_j\rangle$ of the MH excitations in cubic approximation.
In layered LaSrFeO$_4$ with tetragonal crystal symmetry,
three MH peaks are expected in $\sigma_1^a$. They correspond to
(e) an electron transfer between orbitals of the same type, energy $E$,
(d) a transfer from $x^2 \! - \! y^2$ to $3z^2 \! - \! r^2$, energy $E-\Delta_{eg}$, and
(f) a transfer from $3z^2 \! - \! r^2$ to $x^2 \! - \! y^2$, energy $E+\Delta_{eg}$.
}
\label{spins12}
\end{figure}

We now address the MH excitations, which result from the transfer of an electron between neighboring Fe
sites $i$ and $j$ via the $\sigma$ or $\pi$ bonding of the ligand O $2p$ orbital,
% $|$Fe $3d^5_i$ Fe $3d^5_j\rangle \rightarrow$ $|$Fe $3d^4_i$ Fe $3d^6_j\rangle$.
$|3d^5_i \, 3d^5_j\rangle \, \rightarrow \, |3d^4_i \, 3d^6_j\rangle$.
Starting from the cubic approximation, the initial $3d^5$ state has $^6A_1$ symmetry,
corresponding to the $(2S+1)=6$\,-\,fold degenerate
$t_{2g}^3e_g^2$ high-spin state (see Fig.\ \ref{spins12} (a)).
According to the orbital selection rule, there is no overlap between $t_{2g}$ and $e_g$ orbitals on neighboring
sites due to the undistorted 180$^\circ$ bonds in LaSrFeO$_4$.
From the [$^6A_{1g}(d^5); \, ^6A_{1g}(d^5)$] initial state one can reach the final states
[$^5T_{2g}(t_{2g}^2,e_g^2); \, ^5T_{2g}(t_{2g}^4,e_g^2)$]
(corresponding to an electron transfer between $t_{2g}$ orbitals, see Fig.\ \ref{spins12}(b))
and [$^5E_g(t_{2g}^3,e_g^1); \, ^5E_g(t_{2g}^3,e_g^3)$]
(transfer between $e_g$ orbitals, see Fig.\ \ref{spins12}(c)).
All reachable $3d^4$ and $3d^6$ states have total spin $S=2$.
The spectral weight of these transitions therefore strongly depends on the nearest-neighbor
spin-spin correlations,\cite{khaliullin2004a,khaliullinrev,oles2005}
favoring antiparallel alignment of spins on neighboring sites (see Fig.\ \ref{spins12}),
as given in the antiferromagnetically ordered state below $T_N = 366$\,K.\@
Since the orbital quantum number is preserved in the transition
(cf.\ Figs.\ \ref{spins12}b) and \ref{spins12}c)),
both excitations have approximately the same energy.\cite{10Dq}
As a consequence, only one MH peak is expected in the optical spectra in the cubic approximation.

However, deviations from cubic symmetry play an important role in the tetragonal structure of LaSrFeO$_4$.
First of all, the matrix elements for MH excitations between different FeO$_2$ layers can be neglected,
thus MH excitations do not contribute to $\sigma_1^c(\omega)$.
Second, lifting the degeneracy in particular of the $e_g$ level enhances the number of observable absorption
bands in $\sigma_1^a(\omega)$.
An $e_g$ splitting $\Delta_{eg}=\Delta[(x^2 \! - \! y^2)-(3z^2 \! - \! r^2)]$ on the order of 1\,eV
is expected due to the elongation of the FeO$_6$ octahedra,
as discussed in Sec.\ \ref{struc}.
In the following, we neglect the much smaller spitting $\Delta_{t2g}$ within the $t_{2g}$ levels.
We will show that $\Delta_{eg}$ is crucial in order to pull the lowest MH absorption band
below the onset of CT excitations.

\begin{table}[tb]
\caption{Matrix elements for MH excitations between nearest-neighbor Fe sites along
the $x$ direction as given by the Slater-Koster table.\cite{harrison}
% of interatomic matrix elements.\cite{harrison}
The electron transfer takes place via $\sigma$ or $\pi$ bonding of the ligand O $2p$ orbital.
The orbital character changes only for excitations from $x^2\!-\!y^2$ to $3z^2\!-\!r^2$ or vice
versa (third column), in all other cases the orbital character is preserved.
According to Ref.\ [\onlinecite{harrison}], the relation $V_{pd\sigma}=V_{pd\pi} \cdot \sqrt{3}$ holds true.
\label{tab:Harrison}
}
\begin{ruledtabular}
\begin{tabular}{lccccr}
 $x^2\!-\!y^2$ & $3z^2\!-\!r^2$ & $x^2\!-\!y^2 / 3z^2\!-\!r^2$ & $xy$ & $yz$ &  $zx$\\[0.1em]
 \hline\\
  $\frac{3}{4}V_{pd\sigma}^2$ & $\frac{1}{4}V_{pd\sigma}^2$ & $-\frac{\sqrt{3}}{4}V_{pd\sigma}^2$ & $V_{pd\pi}^2$ & 0 & $V_{pd\pi}^2$ \\
  \end{tabular}
  \end{ruledtabular}
\end{table}

The matrix elements for nearest-neighbor Fe - Fe transitions are summarized in Table \ref{tab:Harrison}.
Due to the undistorted 180$^\circ$ bonds of LaSrFeO$_4$, these hopping process conserve the orbital character.
The single exception is the finite overlap between $3z^2 \! - \! r^2$ on site $i$ and $x^2 \! - \! y^2$
on a neighboring site. One thus expects three different MH peaks (referred to as MH1, MH2, and MH3 in the following) in
$\sigma_1^a(\omega)$ with energy separation $\Delta_{eg}$:
(1) The electron transfer from $x^2 \! - \! y^2$ to $3z^2 \! - \! r^2$ (MH1, see Fig. \ref{spins12}(d)).
(2) The excitation from any orbital on site $i$ to an orbital of the same type on the neighboring site
(MH2, see Fig.\ \ref{spins12}(e)).
This excitation is expected at an energy of $\Delta_{eg}$ above MH1.
The individual contributions have approximately the same energy because the orbital quantum number is preserved.\cite{10Dq}
(3) The excitation from  $3z^2 \! - \! r^2$ to  $x^2 \! - \! y^2$ (MH3, see Fig.\ \ref{spins12}(f)).
This excitation is expected at an energy of $2\Delta_{eg}$ above MH1.

We use the matrix elements of Table \ref{tab:Harrison} to calculate the relative spectral weight of the MH peaks.
Summing up the individual contributions we find that the spectral weight of MH1 and MH3 is identical,
whereas the spectral weight of MH2 is expected to be 3.8 times larger.

\begin{figure}[tb]
\includegraphics[width=0.88\columnwidth,clip]{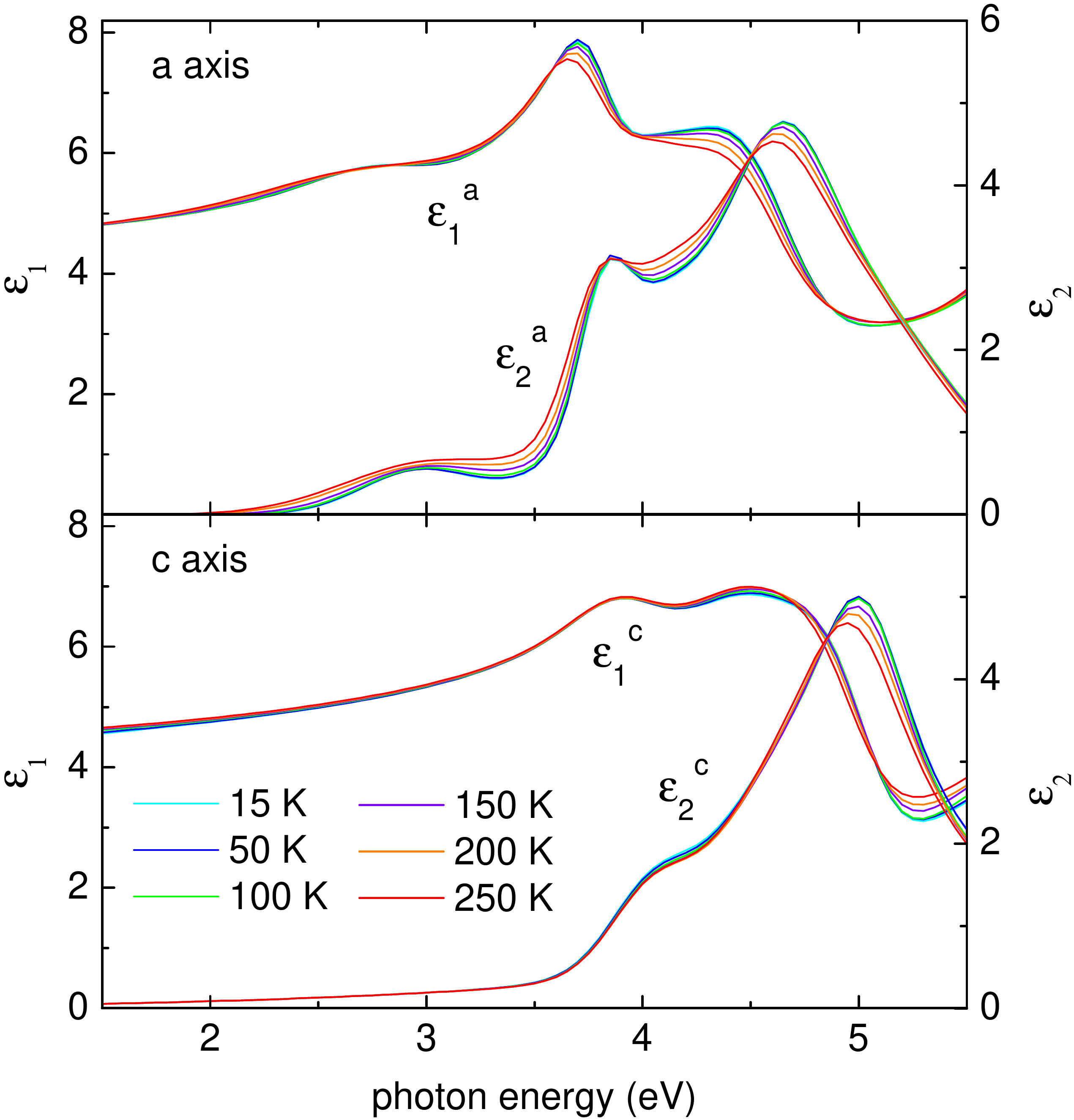}
\caption{(Color online) Dielectric function of LaSrFeO$_4$ for the $a$ and $c$ direction for temperatures between 15\,K and 250\,K. }
\label{eps12LaSrFeO4}
\end{figure}

\section{Results}
\label{results}
\subsection {Experimental data}
\label{expdata}

Figure \ref {eps12LaSrFeO4} displays the dielectric function $\varepsilon(\omega)=\varepsilon_1^l+i\varepsilon_2^l $ ($l=a, c$)
from 1.5\,eV to 5.5\,eV as obtained from the ellipsometric measurements. The real part of the optical conductivity
$\sigma_1^l(\omega)=(\omega/4\pi)\cdot \varepsilon_2^l(\omega)$ is shown in Fig.\ \ref {sig1LaSrFeO4}.
Overall, our data agree with the room-temperature data of $\sigma_1^a(\omega)$ reported in Refs.\ [\onlinecite{moritomo1995,tajima}].
Thus far, data for $\sigma_1^c(\omega)$ has not been reported. We find a striking anisotropy.
Both $\sigma_1^a(\omega)$ and $\sigma_1^c(\omega)$ show a strong absorption band at 4 - 5.5\,eV.\@
However, in $\sigma_1^a(\omega)$ we find an additional peak at 3\,eV
and the shoulder at 3.8\,eV is much more pronounced.
In the following we argue that the latter two features correspond to MH excitations.

\subsection{Peak assignment}
\label{peakassignment}

The observed absorption bands can be assigned to MH and CT excitations.
The role played by local crystal-field (i.e., valence-conserving $3d^5 \rightarrow 3d^{5,*}$) excitations can be neglected
in our analysis of the ellipsometry data.
In LaSrFeO$_4$ they are forbidden both by parity and by the spin selection rule.
Accordingly, they show a very small spectral weight with typical values\cite{rueckamp,figgis} of $\sigma_1 < 10\,(\Omega$cm$)^{-1}$.
Their signatures are visible in the transmittance data below the charge gap, see lower panel of Fig.\ \ref{pseudo}.

\begin{figure}[tb]
\includegraphics[width=0.86\columnwidth,clip]{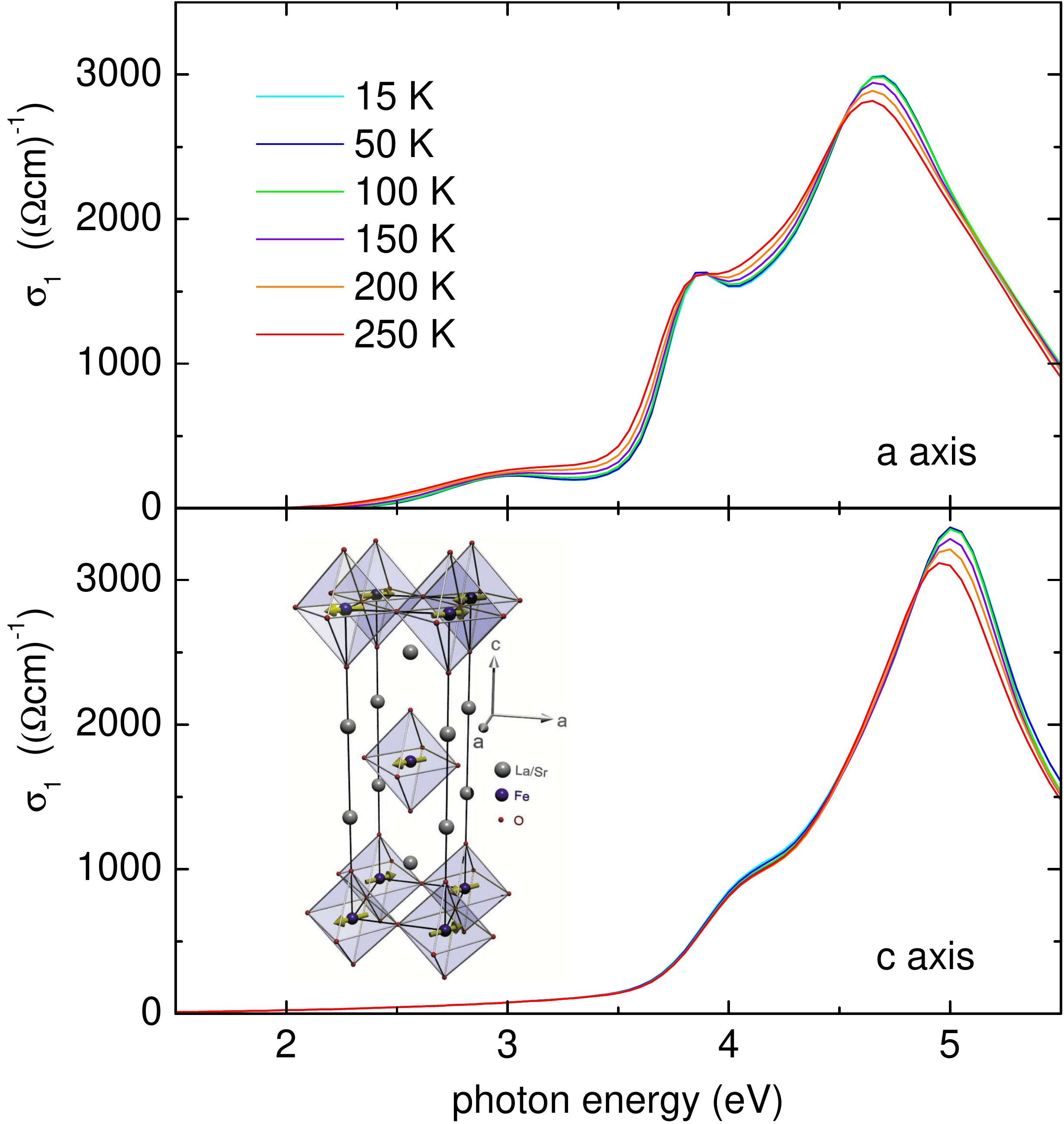}
\caption{(Color online) Optical conductivity $\sigma_1$ of LaSrFeO$_4$ for the $a$ and $c$ direction
for temperatures between 15\,K and 250\,K.\@
Inset: crystal and magnetic structure.\cite{qureshi}  }
\label{sig1LaSrFeO4}
\end{figure}

A first tool to distinguish CT and MH excitations is the spectral weight.
In transition-metal oxides, typical values of $\sigma_1(\omega)$ are of a few 1000\,$(\Omega$cm$)^{-1}$ for CT excitations
but only a few 100\,$(\Omega$cm$)^{-1}$ for MH excitations.\cite{arima1993,goesslingTi,goessling08,reul}
The difference is due to the fact that the matrix elements are of first order in the Fe - O hopping amplitude
for CT excitations and of second order for MH excitations.
To disentangle CT excitations and MH excitations, we further make use of the observed anisotropy.
In layered LaSrFeO$_4$, MH excitations do not contribute to $\sigma_1^c(\omega)$
since the interlayer Fe - Fe hopping is strongly suppressed.

\begin{figure}[tb]
\includegraphics[width=0.8\columnwidth,clip]{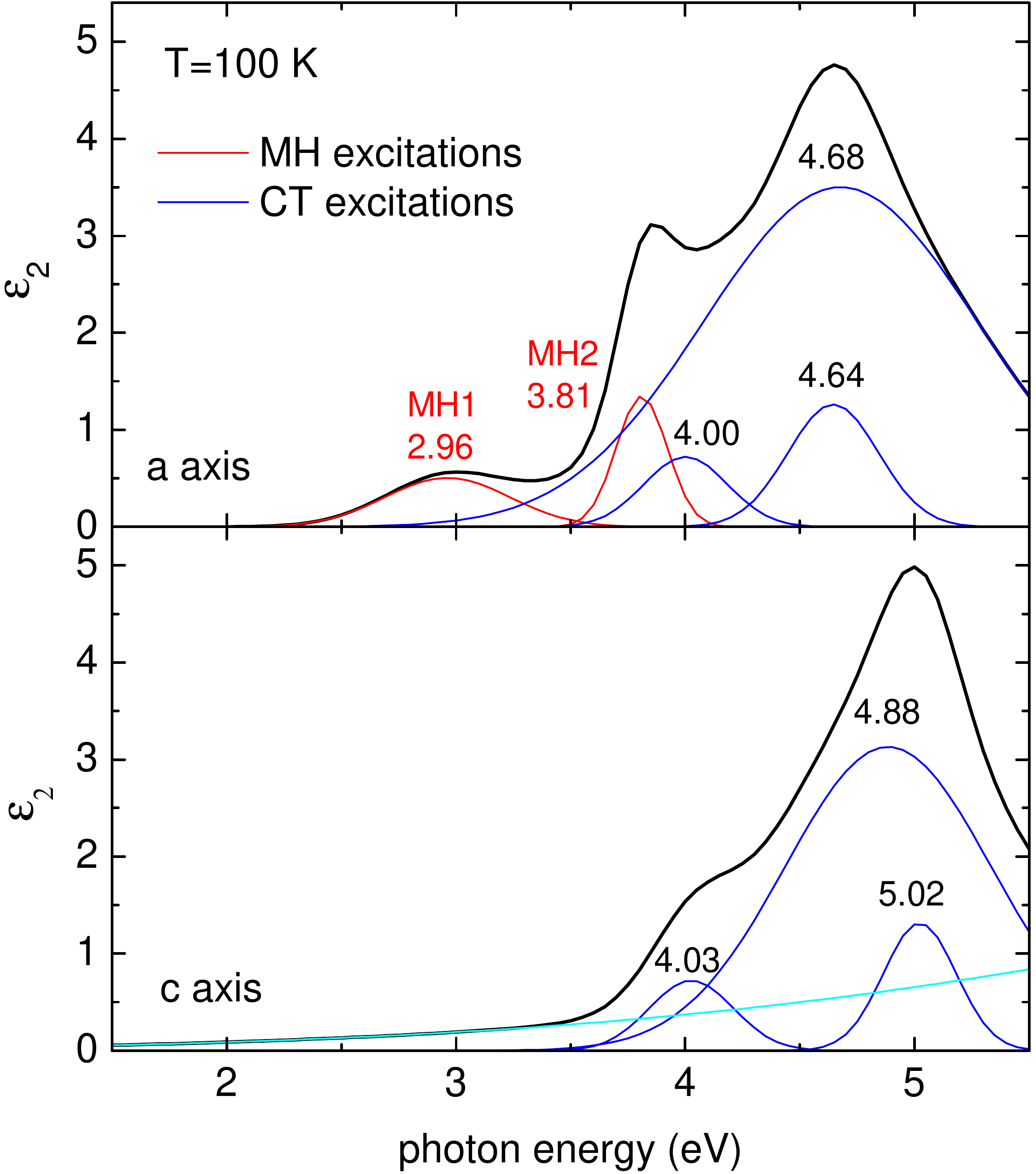}
\caption{(Color online) The measured data is best fit by a sum of five Gaussian oscillators for $\varepsilon_2^a$
and three Gaussian oscillators for $\varepsilon_2^c$, plus an additional oscillator (light blue) outside the measured range
that accounts for higher lying excitations as well as for the small values of absorption below 3.5\,eV in $\varepsilon_2^c$.   }
\label{oszlasrfeo4}
\end{figure}

The strong absorption band observed at 4 - 5\,eV in $\sigma_1^c(\omega)$ clearly has to be attributed to CT excitations.
Our analysis of the ellipsometric data uses three Gaussian oscillators to describe the line shape,
see Fig.\ \ref{oszlasrfeo4}. This does not imply the existence of three microscopically different excitations
since the line shape of the absorption band is not necessarily Gaussian but reflects bandstructure effects.
The analysis of the $a$-axis data requires three very similar oscillators, in particular with a similar spectral weight.
The peak energies differ by up to 0.4\,eV for the two crystallographic directions, which most probably reflects the different
on-site energies of apical and in-plane O ions. We attribute this band at 4 - 5\,eV to the two strong dipole-allowed CT excitations
$t_{2u}(\pi)\rightarrow t_{2g}(\pi)$ and $t_{1u}(\pi)\rightarrow t_{2g}(\pi)$ (in cubic approximation, see discussion
in Sec.\ \ref{possex}). The splitting between these two excitations is expected to be about 1\,eV according to
quantum-chemistry calculations for LaFeO$_3$.\cite{pisarev09} In LaSrFeO$_4$, the next higher-lying peak is observed
at about 7\,eV in the in-plane data of Refs.\ [\onlinecite{moritomo1995,tajima}]. This large energy difference to the
peak at 5\,eV supports our interpretation that both $t_{2u}(\pi)\rightarrow t_{2g}(\pi)$ and $t_{1u}(\pi)\rightarrow t_{2g}(\pi)$
contribute to the absorption band between 4 and 5\,eV.\@
Note that both excitations correspond to a transfer to a $3d$ $t_{2g}(\pi)$ state, and that the crystal-field splitting
of the $t_{2g}(\pi)$ level is expected to be only small, about 0.2\,eV (see Sec.\ \ref{struc}).
Moreover, the matrix elements for transitions into the $t_{2g}$ manifold do not differ very strongly between $a$ and $c$,
even for an elongated octahedron, in contrast to the matrix elements for transitions into the $x^2 \! - \! y^2$ orbital.
The similar spectral weights along $a$ and $c$ between 4\,eV and 5\,eV therefore support our assignment.

The $a$-axis data show two additional features at 3.0\,eV and 3.8\,eV, see Figs.\ \ref{sig1LaSrFeO4} and \ref{oszlasrfeo4}.
For the feature at 3.0\,eV, both its lower spectral weight with $\sigma_1^a(3$\,eV) peaking at about 250\,$(\Omega$cm$)^{-1}$
and the observed anisotropy support an interpretation in terms of a MH excitation.
Moreover, the splitting between the two lowest MH excitations is expected to be roughly
$\Delta_{eg} \! \approx \! 1$\,eV (see Fig.\ \ref{spins12}),
in very good agreement with the difference of 0.8\,eV between the observed peak energies of 3.0 and 3.8\,eV.\@
The third MH excitation is expected roughly $2\Delta_{eg}$ above the lowest one, i.e., within the strong CT band.
As far as the relative spectral weight is concerned, we roughly expect a factor of 3.8 between the two lowest MH
excitations, see Sec.\ \ref{possex}. Experimentally, the spectral weight of MH1 and MH2 is rather similar.
However, our simple estimate does not take into account hybridization effects and is based on a local approach.

At first sight, it is unexpected that the lowest absorption band is of MH type because the $3d^5$ configuration
is stabilized by the intra-atomic Hund exchange $J_H$.
In cubic approximation, a first rough estimate of the MH excitation energy yields $U \! + \! 4J_H$ with $4J_H \approx 3$\,eV.\@
This is much larger than in the $3d^4$ manganites, for which we expect $U \! - \! J_H$.
However, MH1 in the layered structure of LaSrFeO$_4$ corresponds to a transfer from $x^2 \! - \! y^2$ to $3z^2 \! - \! r^2$,
thus it is $\Delta_{eg}^{Fe}$ lower in energy than in cubic approximation.
In comparison, the lowest MH excitation in LaSrMnO$_4$ requires the opposite transfer from $3z^2 \! - \! r^2$ to $x^2 \! - \! y^2$,
raising the excitation energy to $E_{Mn}\! = \! U \! - \! J_H \! + \! \Delta_{eg}^{Mn}$.
In LaSrMnO$_4$, this MH excitation is observed at $E_{Mn} \! \approx \! 2$\,eV.\cite{goessling08}
We thus expect MH1 in LaSrFeO$_4$ at about $E$(MH1)$ \approx \! E_{Mn} \! +  \! 5J_H \! - \! \Delta_{eg}^{Fe} \! - \! \Delta_{eg}^{Mn}$,
i.e., roughly at 3 - 4\,eV.\@ Here, we neglect the slight increase of $U$ from Mn to Fe, but we also neglect
that the 3d$^4$ and 3d$^6$ states relevant for the MH1 excitation both are Jahn-Teller active (cf.\ Fig.\ \ref{spins12}d)),
which reduces $E$(MH1).
Moreover, these estimates neglect the effect of hybridization depicted schematically in Fig.\ \ref{MHCT}.
Therefore, the assignment of the peak at 3.0\,eV to MH1 appears feasible.
% The sizeable $e_g$ splitting in the layered structure is crucial for the reduction of the excitation energy.

However, we also have to discuss alternative scenarios.
As discussed in Sec.\ \ref{possex}, the lowest CT excitation $t_{1g}(\pi)\rightarrow t_{2g}(\pi)$ is dipole-forbidden
and expected at about 0.8\,eV below the lowest dipole-allowed CT excitation. Firstly, the spectral weight of the peak
at 3\,eV is too large for a dipole-forbidden excitation, and secondly, we expect only a modest anisotropy of this
excitation. Possibly, this dipole-forbidden excitation may explain the small but finite values of $\sigma_1^c(\omega)$
between 2\,eV and 3.5\,eV.\@
In a further scenario, the peak at 3.0\,eV may be interpreted as a CT exciton. Note that this peak is lying at about 0.5\,eV
below the CT absorption edge and that a truly bound state with such a large binding energy is very unlikely.
Again, it is not obvious why such an exciton should show a pronounced anisotropy. Moreover, an exciton with
such a large binding energy is expected to show a larger spectral weight and a smaller line width.

Summarizing this section, we have found strong evidence that the lowest dipole-allowed absorption band in LaSrFeO$_4$
is of MH type, which is made possible by the strong splitting $\Delta_{eg}$ caused by the layered structure
and by the Fe $3d$ - O $2p$ hybridization.
We want to add that in LaSrFeO$_4$ the energy of the lowest \textit{dipole-forbidden} CT excitation may be comparable
to the energy of MH1.
Furthermore, the MH excitation from $x^2\! - \! y^2$ at site $i$ to a $t_{2g}$ orbital on a neighboring site
is lower in energy than MH1, but the matrix element for this excitation vanishes (cf.\ Table I).
We emphasize that our results do not disagree with the common interpretation
that non-layered ferrites belong to the class of CT insulators. The different character can be explained by the
absence of a large $\Delta_{eg}$ in the non-layered compounds.
Pisarev \textit{et al.}\cite{pisarev09} studied the optical properties of a series of different ferrites with
trivalent Fe ions. Many of these compounds show a shoulder in the vicinity of the onset of strong CT absorption,
which has been attributed\cite{pisarev09} to the parity-forbidden excitation $t_{1g}(\pi)\rightarrow t_{2g}(\pi)$.
The peak observed at 3.0\,eV in LaSrFeO$_4$ is much too strong for such a dipole-forbidden excitation.
However, our results suggest that MH excitations may not be neglected for a quantitative analysis of the non-layered
ferrites, they may for instance provide a better explanation for a shoulder close to the absorption edge than the
dipole-forbidden excitation discussed above.

\subsection{Temperature dependence}
\label{tdep}

As discussed in the introduction, the temperature dependence of the spectral weight of MH excitations
has attracted considerable interest in different transition-metal compounds because it allows to
study the temperature dependence of nearest-neighbor spin-spin and orbital-orbital
correlations.\cite{khaliullin2004a,khaliullinrev,oles2005,kovaleva2004a,goessling08,lee2005a,fang2003,goesslingTi,reul}
In LaSrFeO$_4$ we expect that the temperature dependence of these correlations is only small below 300\,K.\@
Actually, the spin-spin correlations change only gradually even above the ordering temperature due to the
two-dimensional character.\cite{goessling08} This ferrite thus may serve as a reference compound to study the role of
other effects such as the thermal expansion of the lattice or bandstructure effects.

\begin{figure}[tb]
\includegraphics[width=0.8\columnwidth,clip]{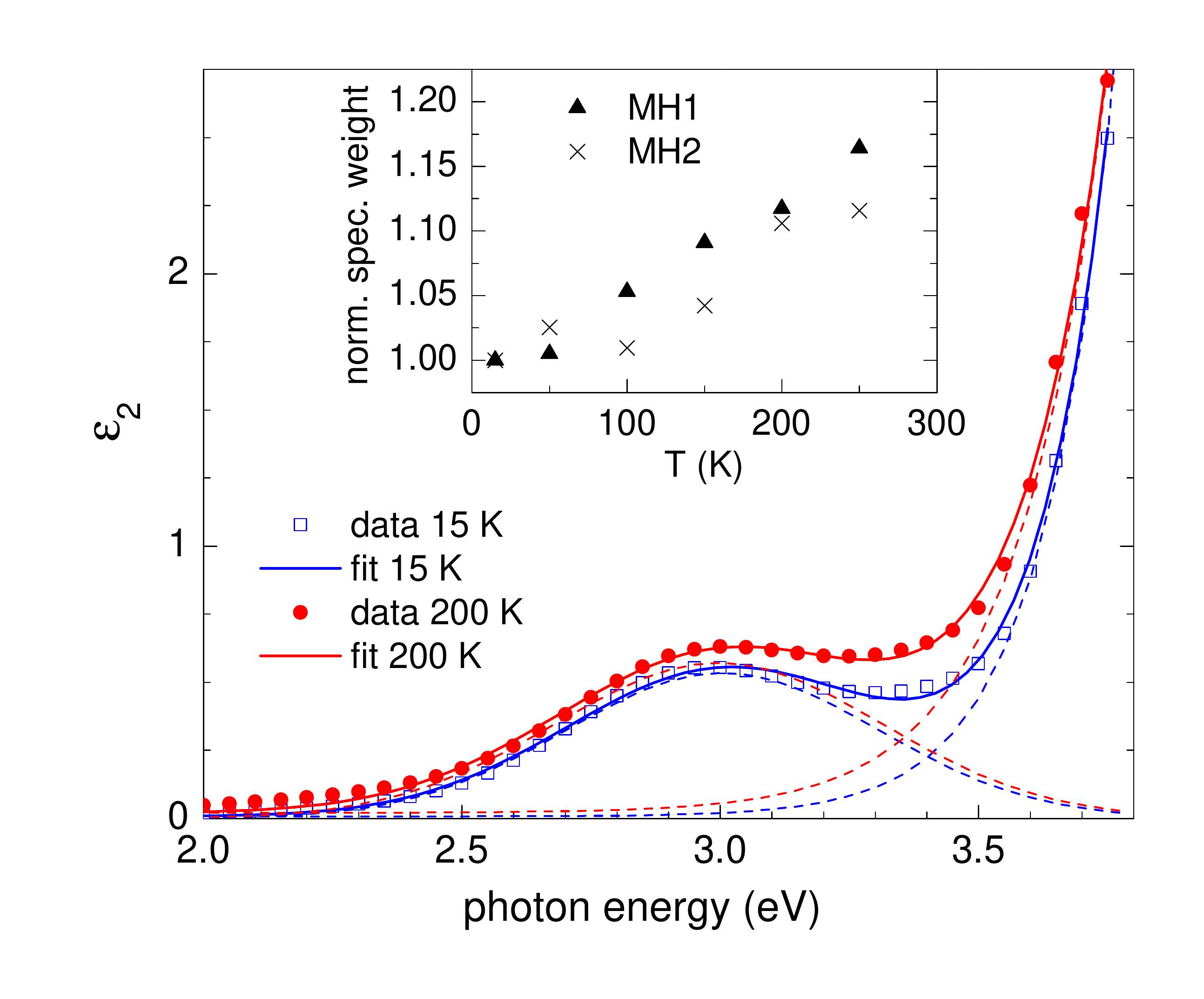}
\caption{(Color online) Solid lines: Fit of the excitation MH1 at 3.0\,eV and of the CT band edge by using a
Gaussian oscillator and an exponential function (depicted by dashed lines).
The temperature dependence clearly is dominated by the change of the band edge.
Inset: Temperature dependence of the normalized spectral weight of MH1 and MH2.
}
\label{fitgausexp}
\end{figure}

As expected, the MH excitations at 3.0\,eV and 3.8\,eV show only a modest temperature dependence,
see top panel of Fig.\ \ref{sig1LaSrFeO4}.
According to the fit using Gaussian oscillators (cf.\ Fig.\ \ref{oszlasrfeo4}),
the spectral weight of MH2 at 3.8\,eV changes only by about 10\,\% between 5\,K and 250\,K, see inset of Fig.\ \ref{fitgausexp}.
At the same time, the peak width increases by about 15\,\%, and the peak frequency decreases by about 1\,\%.
For the peak MH1 at 3.0\,eV, a quantitative analysis is more challenging. Both the spectral weight and the width
of the Gaussian oscillator depicted in Fig.\ \ref{oszlasrfeo4} increase strongly with temperature,
while the frequency of the oscillator \textit{increases} by about 1\,\% from 5\,K to 250\,K.\@
As mentioned above, there is not necessarily a one-to-one correspondence between the Gaussian oscillators
and the microscopic excitations with different line shapes, which is corroborated by the unexpected behavior
of the oscillator parameters such as the hardening of the frequency with increasing temperature.
This gives clear evidence that the change of MH1 is covered by the temperature-induced smearing of the much stronger CT excitations.
To determine the temperature dependence of MH1 more reliably, we separate MH1 from the higher-lying excitations
by fitting $\varepsilon_2^a$ in the range 0.75 - 3.75\,eV simultaneously by an exponential function and a Gaussian
oscillator (see Fig.\ \ref{fitgausexp}). The former accounts for the CT band edge and its shift with temperature,
the latter describes the remaining spectral weight below the CT edge. With this procedure we find an increase
of the spectral weight of MH1 of merely 15\,\% between 15\,K and 250\,K, see inset of Fig.\ \ref{fitgausexp}.
Obviously, also this value has to be taken with care, since it depends strongly on the line shape assumed
for the onset of the CT absorption band.
In fact, we expect the opposite trend, namely
% Lattice effects, on the contrary, should result in
a reduction of spectral weight with increasing temperature
as the $a$ axis lattice constant increases from 3.8709(1) \AA \ to 3.8744(1) \AA \ between 10\,K and room
temperature.\cite{qureshi} An increased Fe - O distance should result in a reduction of the
Fe $3d$ - O $2p$ overlap which in turn should reduce the spectral weight of both MH and CT excitations.

We have to conclude that an accurate determination of the temperature dependence of the spectral weight
of the MH excitations is a difficult task in LaSrFeO$_4$ due to the overlap with the much stronger CT excitations.
A clear separation of strong CT excitations and weaker MH excitations is an obvious prerequisite in order to
reliably determine the spectral weight of the latter.
However, it can safely be concluded that the thermal expansion of the lattice has only a modest impact on
the spectral weight of MH excitations below room temperature.

\section{Summary and conclusions}
\label{sum}

We present a detailed analysis of the optical conductivity of layered LaSrFeO$_4$ for temperatures
ranging from 15\,K to 250\,K in a broad frequency range from 0.5\,eV to 5.5\,eV.\@
% by the use of spectroscopic ellipsometry and transmittance measurements.
Both the anisotropy and the different spectral weight allow us to disentangle Mott-Hubbard and charge-transfer
excitations. We arrive at a consistent assignment of all absorption bands and find strong evidence that the lowest
dipole-allowed excitation is of Mott-Hubbard type. Remarkably, this result is in agreement with previous
studies of non-layered ferrites which have been identified as charge-transfer insulators.
The difference can be explained by the large splitting of the $e_g$ orbital in the layered structure,
which partially compensates the large intra-atomic exchange contribution and pulls the lowest Mott-Hubbard excitation
below the onset of charge-transfer excitations.
\\

\section*{Acknowledgement}

It is a pleasure to acknowledge fruitful discussions with D.I. Khomskii.
This work was supported by the DFG via SFB 608 and BCGS.


\begin{thebibliography}{99}
\bibitem{10Dq}
Here we neglect that the value of 10\,Dq(3d$^n$) depends on the number $n$ of electrons.
\end{thebibliography}
\end{document}